\documentclass{article}

\usepackage{arxiv}

\usepackage[utf8]{inputenc} 
\usepackage[T1]{fontenc}    
\usepackage{hyperref}       
\usepackage{url}            
\usepackage{booktabs}       
\usepackage{amsfonts}       
\usepackage{nicefrac}       
\usepackage{microtype}      
\usepackage{lipsum}		
\usepackage{graphicx}
\usepackage{natbib}
\usepackage{doi}

\title{High-confidence pseudo-labels for domain adaptation in COVID-19 detection}

\author{ \href{https://orcid.org/0000-0003-1274-6750}{\includegraphics[scale=0.06]{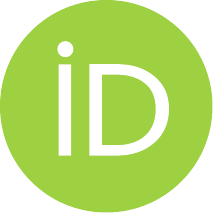}\hspace{1mm}Robert ~Turnbull} \\
	Melbourne Data analytics Platform\\
	The University of Melbourne\\
	Parkville, VIC 3053 \\
	\texttt{robert.turnbull@unimelb.edu.au} \\
	\And
	\href{https://orcid.org/0000-0002-3166-4614}{\includegraphics[scale=0.06]{orcid.pdf}\hspace{1mm}Simon ~Mutch} \\
	Melbourne Data analytics Platform\\
	The University of Melbourne\\
	Parkville, VIC 3053 \\
	\texttt{simon.mutch@unimelb.edu.au} \\
}

\renewcommand{\headeright}{Preprint}
\renewcommand{\undertitle}{Preprint}
\renewcommand{\shorttitle}{High-confidence pseudo-labels}

\hypersetup{
pdftitle={A template for the arxiv style},
pdfauthor={Robert ~Turnbull, Simon ~Mutch},
pdfkeywords={COVID-19, CT Scan},
}

\begin{document}
\maketitle

\begin{abstract}

This paper outlines our submission for the 4th COV19D competition as part of the `Domain adaptation, Explainability, Fairness in AI for Medical Image Analysis' (DEF-AI-MIA) workshop at the Computer Vision and Pattern Recognition Conference (CVPR). The competition consists of two challenges. The first is to train a classifier to detect the presence of COVID-19 from over one thousand CT scans from the COV19-CT-DB database. The second challenge is to perform domain adaptation by taking the dataset from Challenge 1 and adding a small number of scans (some annotated and other not) for a different distribution. We preprocessed the CT scans to segment the lungs, and output volumes with the lungs individually and together. We then trained 3D ResNet and Swin Transformer models on these inputs. We annotated the unlabeled CT scans using an ensemble of these models and chose the high-confidence predictions as pseudo-labels for fine-tuning. This resulted in a best cross-validation mean F1 score of 93.39\% for Challenge 1 and a mean F1 score of 92.15 for Challenge 2.

\end{abstract}


\section{Introduction}

Deep learning models are becoming an increasingly common tool used for medical image analysis. In combination with expert medical professionals, these models can aid in the accurate detection of diseases such as COVID-19 \citep{kollias2020deep, kollias2020transparent}. Here, deep learning models have been shown to provide accurate predictions for the presence of the disease from CT scans alone.

The 4th COV19D competition is being run as part of the `Domain adaptation, Explainability, Fairness in AI for Medical Image Analysis' (DEF-AI-MIA) workshop~\citep{kollias2024domain} at the Computer Vision and Pattern Recognition Conference (CVPR) in 2024. It follows on from previous competitions held as part of the IEEE ICCV 2021~\citep{kollias2021mia}, ECCV 2022~\citep{kollias2022ai} and ICASSP 2023~\citep{kollias2023ai,arsenos2023data} workshops. In the 2024 competition, two challenges presented to participants. The first is to take over one thousand CT scans from the COV19-CT-DB database~\citep{kollias2023deep,arsenos2022large}, annotated as belonging to patients with or without COVID, and train a classifier. The second challenge is to perform domain adaptation. A smaller dataset with CT scans from a different distribution to Challenge 1 is provided. This also includes almost 500 scans which have not been annotated. The challenge is to use the dataset for challenge 1 and make the best classifications on data from a distribution like the additional dataset.

In our submission, we build on work for previous years \citep{cov3d,lungSegmentation2023} where we trained 3D ResNet and SwinTransformer models. In our 2023 submission, we segmented the lungs and cropped the CT scans accordingly. Here we experiment with segmenting both lungs and training additional models with the individual lungs as input. We also use pseudo-labels for augmenting the annotated dataset in the domain adaptation challenge.

\section{Dataset}

The 2024 competition dataset is divided between the two challenges. The Challenge 1 dataset comprises a total of 3,107 scans, with 1,684 used for training and validation (table \ref{tab:challenge1}). We divided the training dataset into four partitions which together with the official validation set gives five partitions for cross-validation. The Challenge 2 specific dataset comprises 4,979 scans, including 912 scans to be used in training and validation. Of these, 494 scans are not labeled as to whether or not the subject is infected with COVID-19. We combined both training and validation partitions from the Challenge 2 dataset and then divided this into roughly equal partitions for five-fold cross-validation.

\begin{table}
  \centering
  \begin{tabular}{cccc}
    \toprule
           & COVID & NON-COVID & Total \\
    \midrule
Training   & 703   & 655       & 1358  \\
Validation & 170   & 156       & 326   \\
Test       & ---     & ---         & 1,413 \\
    \bottomrule
  \end{tabular}
  \caption{The Challenge 1 Dataset.}
  \label{tab:challenge1}
\end{table}

\begin{table}
  \centering
  \begin{tabular}{cccc}
    \toprule
           & COVID & NON-COVID & Total \\
    \midrule
Training    & 120   & 120       & 240   \\
Validation  & 65    & 113       & 178   \\
Unannotated & ---   & ---       & 494   \\
Test        & ---   & ---       & 4,055 \\
    \bottomrule
  \end{tabular}
  \caption{The Challenge 2 Dataset.}
  \label{tab:challenge2}
\end{table}

We also included the public STOIC dataset \citep{stoic} which includes 2,000 CT scans labeled as COVID-19 positive or negative. We ignored the severity categories of COVID-19 positive scans.

\section{Methods}

\subsection{Preprocessing}

As in \cite{lungSegmentation2023}, we first preprocessed the CT scans to segment the lungs and to crop the volumes to the lungs. We then further crop each lung individually. Our aim is not to perfectly segment the lungs from all surrounding tissue, but instead to find the maximum crop that guarantees each lung will be fully contained and contamination from the opposite lung is minimized.
This is achieved by taking each slice in turn, applying a binary threshold using Otsu's method, identifying all contours in the resulting image, removing contours with enclosed areas below a 500 pixels$^2$ and which are clearly not associated with lungs (e.g. span the entire width of the slice), and finally taking the two largest contours which overlap by less than 20\% of their horizontal axis extent. Once we have identified the lungs in each slice, we determine their axis-aligned bounding boxes. The left (/right) lung is cropped to be from the left (/right) side of the volume, to the left (/right) edge of the largest bounding box surrounding the right (/left) lung. A volume containing each lung is stored and these are able to be used as input to the model.

The cropped volumes are interpolated to a single size. The cropped volumes of both lungs are interpolated to $256 \times 256 \times 176$. The individual lungs are interpolated to a size of $320 \times 160 \times 224$.

\subsection{Models}

We use two neural network architectures. The first is a 3D ResNet ~\citep{resnet,Tran2018} with adaptations discussed in \cite{lungSegmentation2023}. The second is a 3D Swin Transformer ~\citep{liu2021}. We used the `Tiny' size of the Swin Transformer to allow for larger CT scan volumes on the GPU. Both neural network models were pretrained on the Kinetics 400 video classification dataset \cite{Kinetics}.

\subsection{Training Procedure}

The models were trained for 30 epochs with a batch size of 2 using cross-entropy loss with the Adam optimizer \citep{adam}. Each volume was included in the training and validation datasets twice with the second one reflected through the sagittal plane. The brightness and contrast for each scan was randomly adjusted during the training according to the scheme discussed in \cite{lungSegmentation2023}.

\subsection{Pseudo-Labels}

For Challenge 2, we train an ensemble of models on the annotated scans and then make predictions on the unannotated scans. These predictions can be used as pseudo-labels \citep{lee2013pseudo}. To mitigate against training with too many scans with incorrect pseudo-labels, we only include predictions with higher confidence, meaning that only include predictions with a probability of the 0.7 or greater. These scans with their pseudo-labels are then included in the training dataset for fine-tuning the models for an additional 10 epochs.

\section{Results}

\subsection{Challenge 1}

\begin{figure}
\centering
\includegraphics[width=\linewidth]{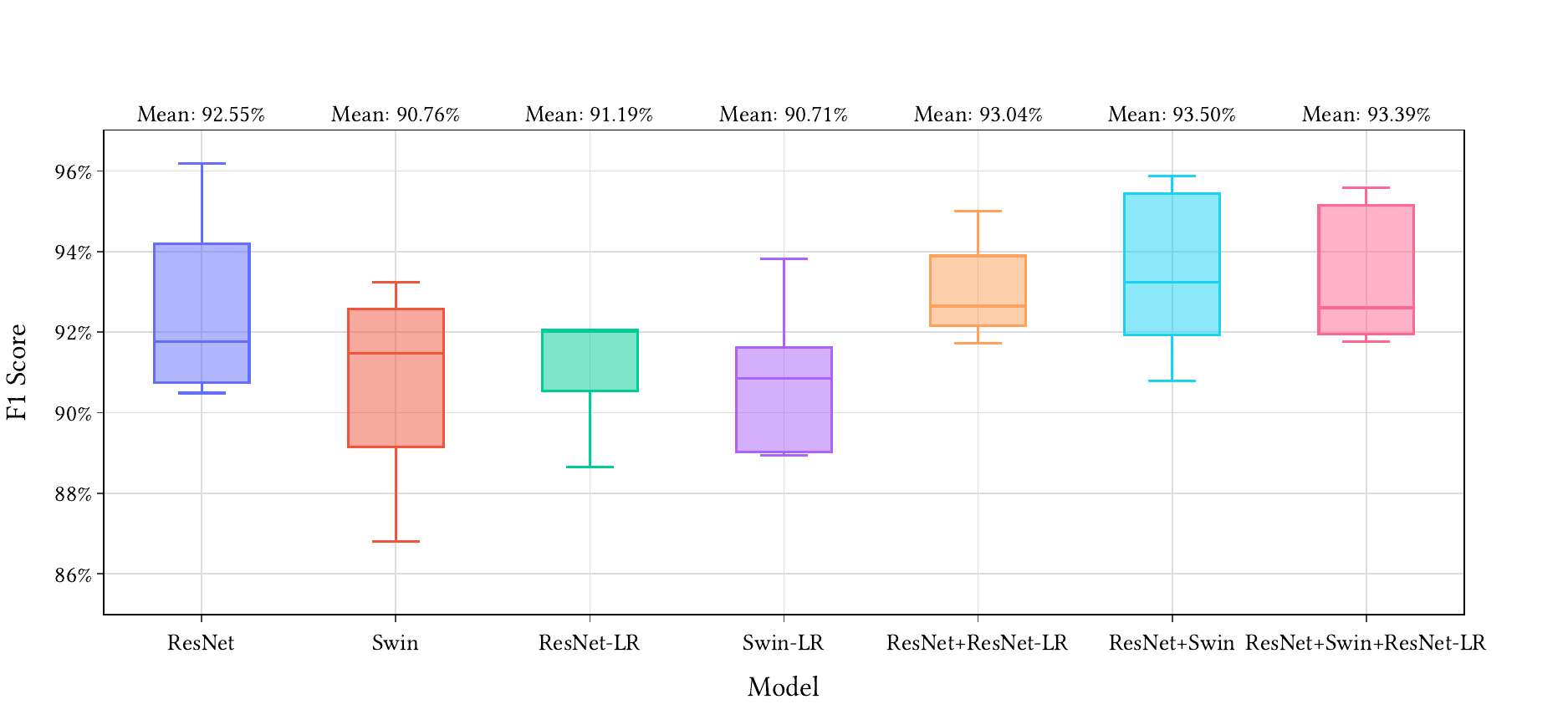}
\caption{The cross-validation results for challenge 1. Models joined with a `+' are ensembles with prediction probabilities averaged.}
\label{fig:challenge1-f1_score}
\end{figure}

Three models were trained for Challenge 1: a ResNet model, a Swin Transformer model and a ResNet model trained on the individual left and right lungs (ResNet-LR). The best performing model was the ResNet which achieved a mean F1 score of 92.55\% across the five cross-validation partitions (fig. \ref{fig:challenge1-f1_score}. Averaging the ResNet and the Swin Transformer results gave the highest F1 score overall at 93.5\%, although including the ResNet-LR model results in the ensemble produced a slightly lower F1 score of 93.4\% but with a smaller variance. The five competition submissions for Challenge 1 are:

\begin{enumerate}
    \item ResNet
    \item Swin Tranformer
    \item Ensemble of ResNet and ResNet-LR
    \item Ensemble of ResNet and Swin Transformer
    \item Ensemble of ResNet, Swin Transformer and ResNet-LR
\end{enumerate}

All submissions average results across models trained on the five cross-validation partitions.

\subsection{Challenge 2}

\begin{figure}
\centering
\includegraphics[width=\linewidth]{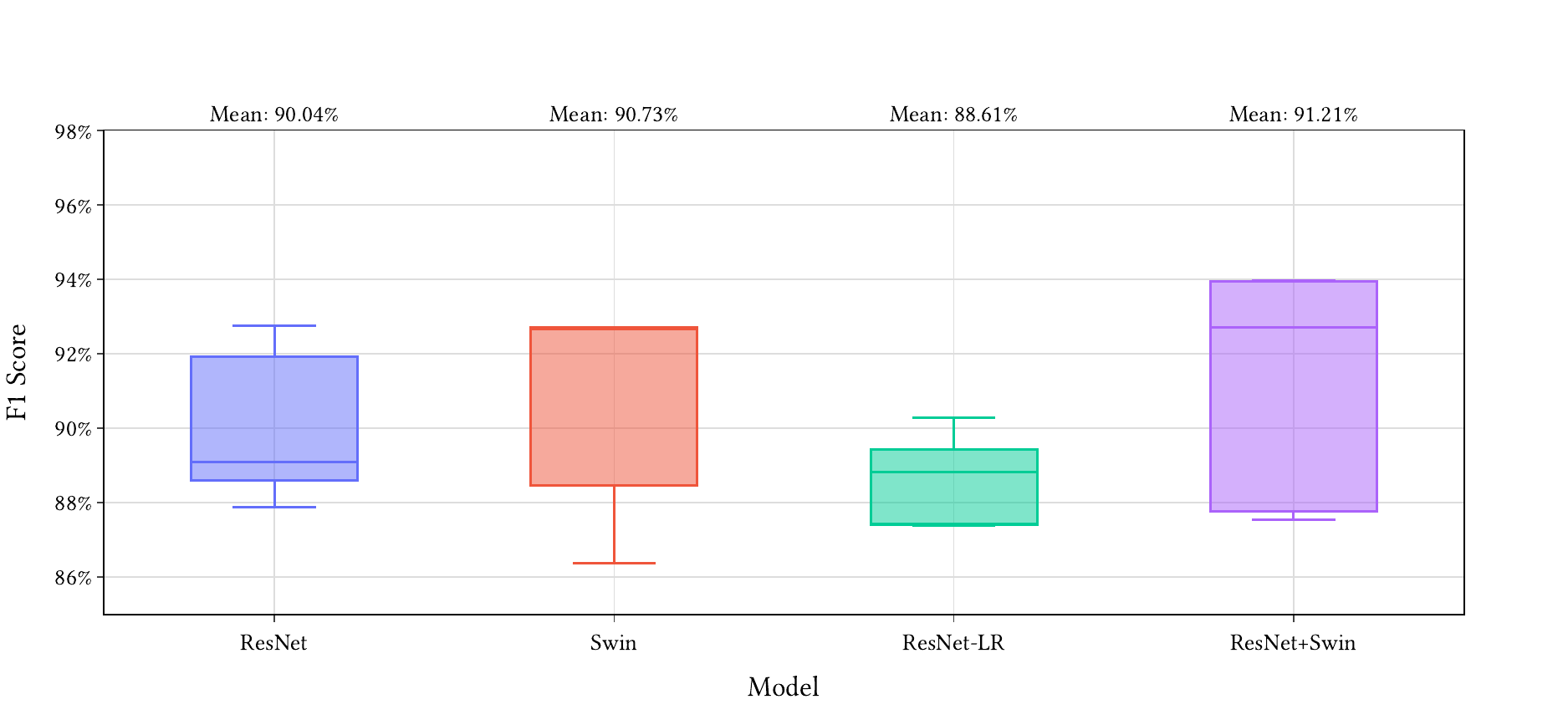}
\caption{The cross-validation results for challenge 2 before adding in pseudo-labels.}
\label{fig:challenge2-f1_score-before-pseudo}
\end{figure}

\begin{figure}
\centering
\includegraphics[width=\linewidth]{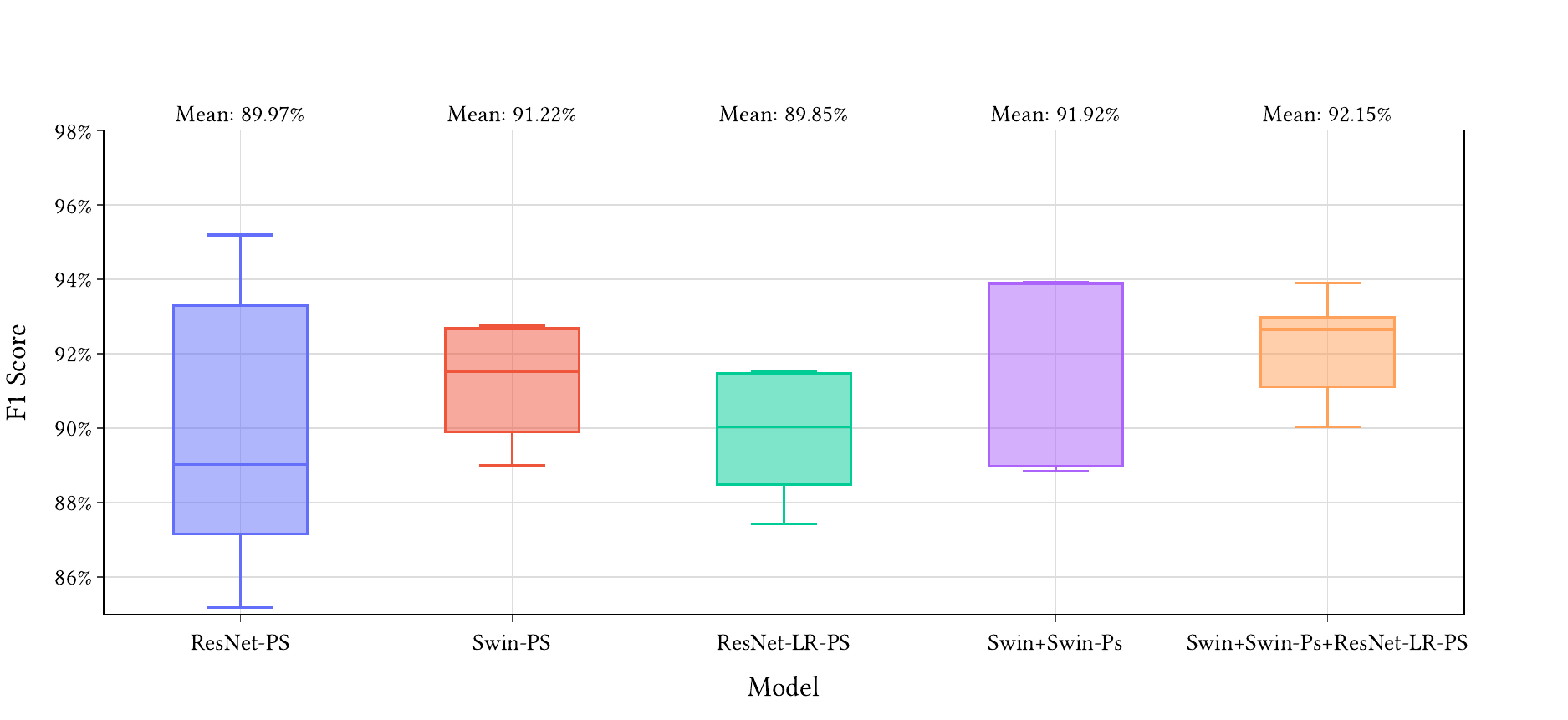}
\caption{The cross-validation results for challenge 2 after adding in pseudo-labels.}
\label{fig:challenge2-f1_score-after-pseudo}
\end{figure}

A ResNet and a Swin Transformer were trained to predict the pseudo-labels. An ensemble of both achieved an F1 score of 91.2\% (fig. \ref{fig:challenge2-f1_score-before-pseudo}). If we filter the validation datasets for only high-confidence scans with a probability of being with or without COVID-19 above 0.7, then the F1 score increases to 95.8\%. Using this ensemble, predictions were made on the 494 unannotated scans for Challenge 2. Of these, 414 predictions were above the threshold of 0.7 and these were assigned as pseudo-labels. This improved the F1 score for the Swin Transformer to 91.22\% but the result for the ResNet decreased a small amount (fig. \ref{fig:challenge2-f1_score-after-pseudo}). The ResNet-LR model improved from 88.6\% to 89.85\%. An ensemble of both Swin Transformer models (with and  without pseudo-labels) together with the ResNet-LR trained with pseudo-labels achieved the highest F1 score of 92.15\%.

The five competition submissions for Challenge 2 are:

\begin{enumerate}
    \item ResNet
    \item Swin Tranformer
    \item Swin Tranformer with pseudo-labels
    \item Ensemble of Swin Transformer and Swin Transformer with pseudo-labels
    \item Ensemble of Swin Transformer and Swin Transformer with pseudo-labels and the ResNet model with individual lungs and pseudo-labels.
\end{enumerate}

\section{Conclusion}

The approach used in this paper achieved high validation F1 scores for both challenges. The best result for Challenge 1 was an ensemble of the ResNet and Swin Tranformer models with an average F1 score of 93.5\%. The best single model for Challenge 2 was the Swin Transformer at an F1 score of 90.73\%. This improved to 91.22\% when pseudo-labels with high-confidence were added to the training set. An ensemble achieved even better results with an F1 score of 92.15\%. These results for the domain adaptation challenge show that high accuracy can be obtained for classification of a new distribution of CT scans with a relatively small number of annotated examples. 

\section{Acknowledgements}
This research was supported by The University of Melbourne’s Research Computing Services and the Petascale Campus. We acknowledge the help of Evelyn Mannix.

\bibliographystyle{unsrtnat}
\bibliography{references}

\end{document}